\documentclass[osajnl,showpacs,twocolumn]{revtex4} 

\usepackage{epsfig,times}

\begin{document}

\title{Photon counting with loop detector}

\author{Konrad Banaszek and Ian A. Walmsley}

\affiliation{Clarendon Laboratory, University of Oxford, Parks Road,
Oxford OX1 3PU, United Kingdom}

\date{\today}

\begin{abstract}
We propose a design for a photon counting detector capable of
resolving multiphoton events. The basic element of the setup
is a fiber loop, which traps the input field with the help of a fast
electrooptic switch. A single weakly coupled avalanche photodiode is used
to detect small portions of the signal field extracted from the loop. We
analyze the response of the loop detector to an arbitrary input field,
and discuss both the reconstruction of the photon number distribution of
an unknown field from the count statistics measured in the setup, and the
application of the detector in conditional state preparation.
\end{abstract}


\maketitle

Photon counting is an important method of detecting light with
applications in many fields including spectroscopy, atmospheric physics,
and quantum information processing. Detection of single photons
necessarily involves a gain mechanism in order to generate a macroscopic
number of photoelectrons from the absorption of a single
energy quantum from the incident electromagnetic radiation. Among the
most popular of detectors with this characteristic are
  avalanche photodiodes operated in the
Geiger mode.\cite{GeigerAPDs} Compared to standard photomultipliers,
they have higher quantum efficiency, are more stable and are more robust
with respect to external conditions.

Geiger-mode avalanche photodiodes have one
drawback: the breakdown current is almost completely independent of the
number of absorbed photons. Consequently, it is
not possible to determine the number of photons incident on the detector
on a time scale short compared to the detector response time.
This has serious implications for several applications. For example,
it critically affects the performance of practical quantum
cryptography systems.\cite{Cryptography} It also impairs the fidelity with
which entangled states may be prepared,
\cite{High NOON} as well as the conditional detection required for linear
optical quantum computing.\cite{KLM}

In this paper, we propose a photon
counting setup which is capable of resolving multiphoton events. The
basic idea is to split the input signal into separate small pieces which
are expected to contain less than one photon, and therefore can be
detected with an avalanche photodiode without losing information on the
photon number. Our method is based on current fiber optic
technology and has the important advantage that it requires only a single
avalanche photodiode. This contrasts with previous proposals that involve
the splitting the signal input, which require a large array of
photodiodes.\cite{KokBrauPRA01} We note that multiphoton absorption
events can be in principle resolved by more exotic photon-counting
detectors, such as solid state photomultipliers.\cite{KimTakeAPL99}
These, however, still remain at the developmental stage, and must be
operated at liquid helium temperatures.

\begin{figure}[b]
\centerline{\epsfig{file=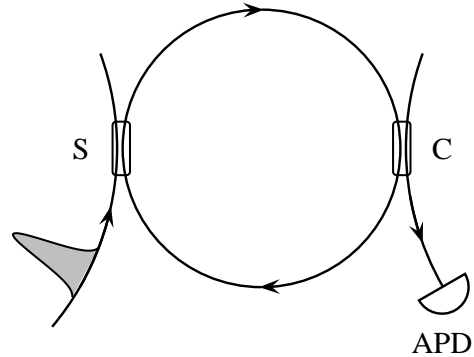,width=2.5in}}
\caption{The proposed construction of the loop
detector. S, electrooptic switch; C, weak coupler; APD, avalanche
photodiode.\label{Fig:Detector}}
\end{figure}

The proposed loop detector is depicted schematically
in Fig.~\ref{Fig:Detector}. The basic element of the setup is a fiber
loop used as a storage cavity. The light pulse whose photon statistics
are to be measured enters the loop through a fast electrooptic switch
S. Immediately after injection of the pulse the state of the switch is
flipped, which closes the loop and thus traps the input pulse. In each
circulation of the loop, a small portion of
the pulse is extracted by a weak static
coupler C and incident on avalanche photodiode
(APD) operated in the Geiger mode. The
duration of a single roundtrip in the loop is longer than
the dead time of the APD, so that after each roundrip the APD is ready for
a new detection event. The quantity recorded in the measurement is the
total number of counts on the photodiode.

The main result of this paper is to show that the statistics of counts on
the APD are related to the photon distribution of the input pulse. We
will first calculate the response of the detector to a coherent field.
This result can be generalized to an arbitrary input field by averaging
with an appropriate probability distribution for the field amplitude. In
the most general case this probability distribution is given by Glauber's
$P$ representation for the input field. Next, we will use this result to
derive a method for reconstructing the photon number distribution from
the count statistics for a completely unknown input field.

We begin by calculating the probability distribution $p_{\cal I}(k)$ of
obtaining $k$ counts on the photodiode assuming that the coherent pulse
after injection into the loop had intensity
${\cal I}$. In the derivation, it will be convenient to use the
generating function:
\begin{equation} \tilde{p}_{\cal I}(z) =
\sum_{k=0}^{\infty} z^k p_{\cal I}(k).
\end{equation}
There are several parameters of the setup relevant to our
calculation. Let $t_r$ be the power transmission of the loop
for a single pulse roundtrip, including losses at all the optical
elements. Further, let $t_c$ be the fraction of the light intensity
extracted by the coupler C to the avalanche photodiode. We assume that the
photodiode is characterized by the quantum efficiency $\eta$, and that
the probability of registering a dark count is $p_d$. We will discuss
realistic values for all these parameters later, though our
analysis is valid for an arbitrary set of parameters satisfying certain
approximations.

The assumption of coherent input substantially simplifies the
calculations, since the portion of the pulse directed to
the APD in each roundtrip is uncorrelated with the
field remaining in the loop. Consequently, the counts registered by the
avalanche photodiode in each roundtrip are statistically independent. In
the $i$th circulation of the loop, the intensity of the pulse is
$t_r^{i-1} {\cal I}$. Of this the light intensity extracted to the
photodiode is $t_c t_r^{i-1} {\cal I}$. Taking the efficiency  of the
avalanche photodiode to be
$\eta$, the probability of a count in the $i$th roundtrip is
$1-(1-p_d)\exp(-\eta t_c t_r^{i-1}{\cal I})$, including the
possibility of a dark count with probability $p_d$.
For an input pulse in a coherent state the counts on the APD are
statistically uncorrelated on each circulation, so that the generating function
$\tilde{p}_{\cal I}(z)$ is simply given by a product of generating 
functions describing
detection events for each round trip:
\begin{equation}
\label{Eq:General}
\tilde{p}_{\cal I}(z) = \prod_{i=1}^{L}
[z + (1-z) (1-p_d) \exp(-\eta t_c t_r^{i-1} {\cal I}) ].
\end{equation}
Here $L$ is the total number of circulations of the loop.

The expression derived in Eq.~(\ref{Eq:General}) describes the most
general situation of arbitrary input intensity. An important limiting case is
when the probability of extracting two or more photons from the
loop in a single roundtrip via the coupler C is negligible. Then
the logarithm of the generating function $\ln \tilde{p}_{\cal I}(z)$ can be
expanded up to terms linear in $t_c {\cal I}$ or $p_d$, assuming that
both these parameters are much smaller than one. Further, when
the number of roundtrips is large enough to allow
all the input signal photons to leak from the loop, we obtain
from the logarithmic expansion the following
approximate expression for the generating function:
\begin{equation}
\tilde{p}_{\cal I}(z) \approx \exp \left[ (z-1) \left(\frac{\eta
t_c {\cal I}}{1-t_r} + L p_d \right) \right].
\end{equation}
This formula describes simply the standard Poissonian statistics 
normally associated
with coherent radiation, with the average number of counts equal to 
$\eta t_c {\cal
I}/(1-t_r) + L p_d$. The first term here describes the average number 
of counts generated
by the input light, and the second term corresponds to dark
counts. In this regime, the response of the loop detector is 
described by a single
effective efficiency parameter:
\begin{equation}
\eta_{\text{eff}} = \eta \frac{t_c}{1-t_r}.
\end{equation}
In the
ideal limit all the optical elements are lossless, and the only source
of attenuation for the pulse trapped inside the loop is extraction of the
photons to the avalanche photodiode. Then $t_c= 1 - t_r$, and the
effective efficiency of the loop detector approaches that of the avalanche
photodiode, but with the advantage of having photon number resolution.
In a realistic case, there are always excess losses in the coupler or
the electroopic switch, which decreases the effective efficiency of the
loop detector below the level of the APD itself.

\begin{figure}
\centerline{\epsfig{file=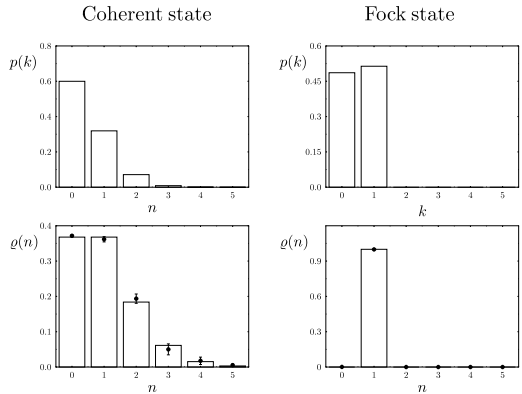,width=3.375in}}

\vspace*{-0.5in}

\caption{Monte Carlo simulation of photon counting using a loop
detector for a coherent state with the average photon number equal to one
(left), and a single-photon Fock state (right). The top graphs depict
simulated count statistics based on $N=10^{5}$ repetitions of
the mesurement, assuming $t_r=72\%$, $t_c=20\%$, $\eta=80\%$, $p_d=0$.
The dots in bottom graphs present the photon number distributions
reconstructed with the help of the singular-value decomposition method,
compared with the exact distributions shown with bars.
\label{Fig:Simulation}}
\end{figure}

We now turn to the problem of reconstructing the photon number
distribution of an unknown input field from the statistics of counts
measured using the loop detector.
Let $\varrho(n)$ be the photon number distribution of the pulse 
immediately after
injection into the fiber loop. The probability $p(k)$ of observing 
$k$ photocounts is
given by the linear combination:
\begin{equation}
\label{Eq:LININPOS}
p(k) = \sum_{n} w(k|n) \varrho(n)
\end{equation}
where $w(k|n)$ are conditional probabilities
of registering $k$ counts on the photodiode given exactly $n$ photons
injected into the loop. These conditional probabilities are
defined exclusively by the parameters of the setup. The explicit form of these
conditional probabilities can be found from our previous calculation 
of the generating
function for a coherent input. For a coherent input pulse, the photon number
distribution is Poissonian,
$\varrho(n) = e^{-{\cal I}} {\cal I}^n/n!$. Inserting this into
Eq.~(\ref{Eq:LININPOS}), multiplying both its sides
by $z^k$ and performing the summation over $k$ yields:
\begin{equation}
e^{\cal I} \tilde{p}_{\cal I}(z)
=
\sum_{k=0}^{\infty} \sum_{n=0}^{\infty} w(k|n)\frac{z^k {\cal I}^n}{n!}
\end{equation}
where $\tilde{p}_{\cal I}(z)$ is defined in Eq.~(\ref{Eq:General}).
Thus the conditional probabilities $w(k|n)$ can be identified from the
generating function
$e^{\cal I} \tilde{p}_{\cal I}(z)$ for a coherent input by
expanding it into a double power series in the parameters ${\cal I}$ and
$z$. This procedure can be easily
performed using any of the standard computer programs for symbolic
algebraic calculations. Given the explicit form of the conditional
probabilities $w(k|n)$ and the measured count statistics $p(k)$, one can then
use one of the standard methods for solving linear systems to
reconstruct the photon number distribution $\varrho(n)$. In
Fig.~\ref{Fig:Simulation} we show the results of a Monte Carlo simulation
of the loop detector operation, using the singular value
decomposition method to reconstruct the photon number distribution. 
The detector is able
to provide very good estimates of the photon number distribution of 
both classical and
nonclassical inputs. The accuracy of the reconstruction can be estimated
using standard statistical tools.\cite{BanaJMO99}

In a number of quantum information processing protocols,\cite{High NOON,KLM}
determination of the photon number has to be made on the single-shot
basis in order to infer non-destructively the state of another quantum
system that is entangled with the detected radiation. In such an application,
the performance of the detector can be characterized by the
confidence:\cite{KokBrauPRA01}
\begin{equation}
C_k = \frac{w(k|k) \varrho(k)}{\sum_{n} w(k|n) \varrho(n)}
\end{equation}
describing {\em a posteriori} probability that a $k$-click event has been
triggered by the $k$-photon component of the input field state. Here
$\varrho(n)$ is the {\em a priori} photon number distribution of the input
field, and $w(k|n)$ are the conditional probabilities calculated before.
In practice, the confidence can be optimized by tuning
the splitting ratio of the coupler C used in the construction of the loop
detector.

An important question is whether realistic values for the parameters
of the loop detector enable it to function in the manner shown in
Fig.~\ref{Fig:Simulation}. In the visible region, silicon avalanche 
photodiodes can
exhibit efficiencies greater 75\% for wavelengths near
700~nm.\cite{DautDescAO93,KwiaSteiPRA93} The dark count rates for 
these devices can be
as low as 25~Hz, which is negligible. This rate may be further reduced by
operating the photodiode in the gated mode.  The APD parameters 
become less favorable
beyound the 1 $\mu$m wavelength region, where germanium or InGaAs materials are
used.\cite{OwenRariAO94,LacaZappAO96,RiboGautAO98} As with single 
photon detection, the
quantum efficiency of the photodiode imposes the upper limit on the 
performance of the
loop detector. A second important factor is the excess losses inside 
the fiber loop,
which attenuate the pulse in each roundtrip in addition to the 
fraction extracted by the
coupler C. These excess losses lead to two opposing constraints on 
the extraction of the
light from the loop to the photodiode, parameterized by the coupling 
losses $t_c$. On
one hand, we have seen that in order to recover the complete photon 
number distribution,
$t_c$ should be as small as possible. On the other hand, if too small a power
is extracted from the loop, then a large number of roundtrips are
required to measure the count statistics. Consequently,
substantial excess losses in the loop would mean that most of the 
photons would disappear
inside the loop before ever reaching the photodiode, and would give 
no signal at all.
The optimal value of $t_c$ is some intermediate value which allows
one to detect a substantial fraction of photons in the input signal 
and at the same time
provides a sufficiently large number of multiphoton events. A good 
test of whether both
these conditions are satisfied is to check the singularity
of the matrix composed of the conditional probabilities $p(k|n)$ for the
range of photon numbers which are expected in the input signal. If the
matrix is not singular, then the count statistics provides sufficient
information to reconstruct the complete input photon number 
distribution from the
experimental data.

In conclusion, we have proposed and analyzed a design for a photon counting
detector capable of resolving multiphoton detection events using 
standard laboratory
technology. The setup uses commonplace fiber-optic components and 
does not require
extreme operating conditions. 

This work was begun when the authors were at The Institute of Optics, 
University of Rochester,
Rochester, New York, 14627, USA.  It was supported by the Center for
Quantum Information, an ARO-administered MURI Center, funded by grant No.
DAAG-19-99-1-0125.  We acknowledge enlightening conversations with Prof.
C. Radzewicz.

\end{document}